\documentclass[conference,a4paper]{IEEEtran}

\usepackage[utf8]{inputenc} 
\usepackage[T1]{fontenc}
\usepackage{url}              
\usepackage{cite}             
\usepackage{xcolor}
\usepackage[cmex10]{amsmath}  
\usepackage{mleftright}       
\mleftright                   

\usepackage{graphicx}         
\usepackage{booktabs}         
\usepackage{arydshln}

\newtheorem{lemma} {Lemma}

\usepackage[justification=justified,skip=0pt]{caption}
\begin{document}
	
	\title{Capacity Bounds for the Poisson-Repeat Channel} 
	
	\author{%
		\IEEEauthorblockN{
		}
		\IEEEauthorblockA{%
		}
	}
	
	%
	%
	%
	 \author{%
		   \IEEEauthorblockN{Mohammad Kazemi and
			                     Tolga~M.~Duman
			   \IEEEauthorblockA{ Department of Electrical and Electronics Engineering\\
				   	 					Bilkent  University, Ankara, Turkey\\
				                     \{kazemi, duman\}@ee.bilkent.edu.tr}
			 }
		 }
	
	
	\maketitle
	
	\begin{abstract}
		We develop bounds on the capacity of Poisson-repeat channels (PRCs) for which each input bit is independently repeated according to a Poisson distribution. The upper bounds are obtained by considering an auxiliary channel where the output lengths corresponding to input blocks of a given length are provided as side information at the receiver. Numerical results show that the resulting upper bounds are significantly tighter than the best known one for a large range of the PRC parameter $\lambda$ (specifically, for $\lambda\ge 0.35$).
		We also describe a way of obtaining capacity lower bounds using information rates of the auxiliary channel and the entropy rate of the provided side information.
	\end{abstract}
	
	\section{Introduction}
	Poisson-repeat channel (PRC) is a discrete memoryless channel (DMC) where each input bit is independently repeated following a Poisson distribution. It was first introduced in \cite{Mitzenmacher2006} as an auxiliary channel to obtain a lower bound on the capacity of the binary deletion channel (BDC). PRC has also been used in some practical scenarios since it can model both deletions (zero repetitions) and more than one repetitions, and it is usually easier to handle than other deletion and repetition models \cite{Mitzenmacher2006}. For instance, it is used to model the errors in single photon generation \cite{Buller2010}, a crucial step in quantum cryptography and key distribution, as well as the errors in distribution support estimation problem \cite{Chien2021}, with applications in estimating the mutational diversity of genes in viral genomes.
	
	There have been some progress on the design of practical codes for the PRC as well \cite{Con2022,Pernice2022,Rubinstein2022}. Explicit construction of a family of error-correcting codes for the BDC with linear-time encoding and decoding is presented in \cite{Con2022}, and it is shown that this construction works for the PRC as well. 
	In \cite{Pernice2022}, the authors first show that a capacity-achieving code for a family of repeat channels including the PRC can be assumed to have an approximate balance in the frequency of zeros and ones of all sufficiently long sub-strings of all codewords. With this assumption, they design explicit codes for a general class of repeat channels with a similar construction as in \cite{Con2022}, with linear-time encoding and quasi-linear-time decoding. A more efficient code construction for the PRC is provided in \cite{Rubinstein2022}, where it is also proven that any family of codes for the PRC can be converted into a family of PRC codes with an arbitrarily close rate that has quasi-linear encoding and decoding complexities. 
	
	Some bounds on the PRC capacity are provided in \cite{Mitzenmacher2006,Cheraghchi2019}. In \cite{Mitzenmacher2006}, it is shown that the BDC capacity can be lower bounded for any deletion probability by lower bounding the PRC capacity. Motivated by this, the authors provide a lower bound on the PRC capacity. Convex programming techniques are employed in \cite{Cheraghchi2019} to obtain capacity upper bounds for a general class of repetition channels including PRC. To the best of our knowledge, these works represent the only available PRC capacity bounds in the literature.
	
	In this paper, we develop bounds on the capacity of PRCs. To this end, we first provide bounds on the capacity of an auxiliary PRC-inspired channel where the output length of each input block of a certain length is provided as side information at the receiver. This way, the PRC breaks into smaller channels with a finite number of inputs; however, the number of possible outputs may still be unbounded due to Poisson-distributed repetitions. To resolve this issue, we prune the output sequences and condition on an event based on the repetition structure, and employ the corresponding results to bound the capacity of the auxiliary channel. Note that since the pruned DMC has a finite number of inputs and outputs, its capacity can be calculated using the Blahut-Arimoto algorithm (BAA). Therefore, it is possible to obtain upper bounds on the PRC capacity. Indeed, the results show that the proposed upper bound improves the best known one significantly. Finally, using the information rates of the auxiliary channel and the entropy rate of the side information, we provide some insight on how lower bounds on the capacity may also be developed.
	
	The rest of the paper is organized as follows. The system model and the auxiliary channels are introduced in Section II. In Section III, we provide upper bounds on the PRC capacity through a simple DMC output partitioning result that suitably prunes the outputs. In order to make the bounds numerically calculable for larger input lengths, in Section IV, we condition on an event based on the repetition structure, and utilize the resulting inequalities on related mutual information terms to develop the upper bounds. Ideas on how the above approach can be used to develop lower bounds on the PRC capacity are presented in Section V. Numerical examples are provided in Section VI, and finally, the paper is concluded in Section VII.
	
	\section{Poisson-Repeat Channel}
	We consider point-to-point communications over a PRC, which is a binary-input binary-output channel with synchronization errors where each input bit is received with repetitions following a Poisson distribution with parameter $\lambda$. Since no repetitions are also possible, the bits may be deleted as well. For instance, over PRC, input `01' can lead to an infinite number of outputs including `1', `00', `01', and `001' with repetition patterns $\left(0,1\right)$, $\left(2,0\right)$, $\left(1,1\right)$, and $\left(2,1\right)$, respectively, where $\left(r_1,r_2\right)$ means that the number of repetitions for the first bit is $r_1$ and that for the second bit is $r_2$.
	
	A lower bound on the capacity of the PRC in terms of a numerically calculable summation is presented in \cite{Mitzenmacher2006} using the jigsaw-puzzle decoding approach developed for the insertion/deletion channels \cite{Drinea2007}. In \cite{Cheraghchi2019}, upper bounds for a general class of repetition channels including PRC are obtained by converting the capacity problem into the problem of maximizing a univariate, real-valued, and often concave function over a bounded interval.
	
	\section{Genie-Aided Upper Bounds on the PRC Capacity}
	In order to obtain calculable bounds on the capacity of a PRC, we first consider a PRC-inspired auxiliary channel by dividing the input sequence of length $N$ into $Q$ blocks of length $L$ bits, and providing the receiver with the length of each received block as side information. We denote this side information as $V$ with $V_i$ being the side information corresponding to the $i$-th block. 
	\subsection{Bounds on the Capacity of the Auxiliary Channel}
	\begin{figure}
		\centering
		\includegraphics[width=.7\linewidth]{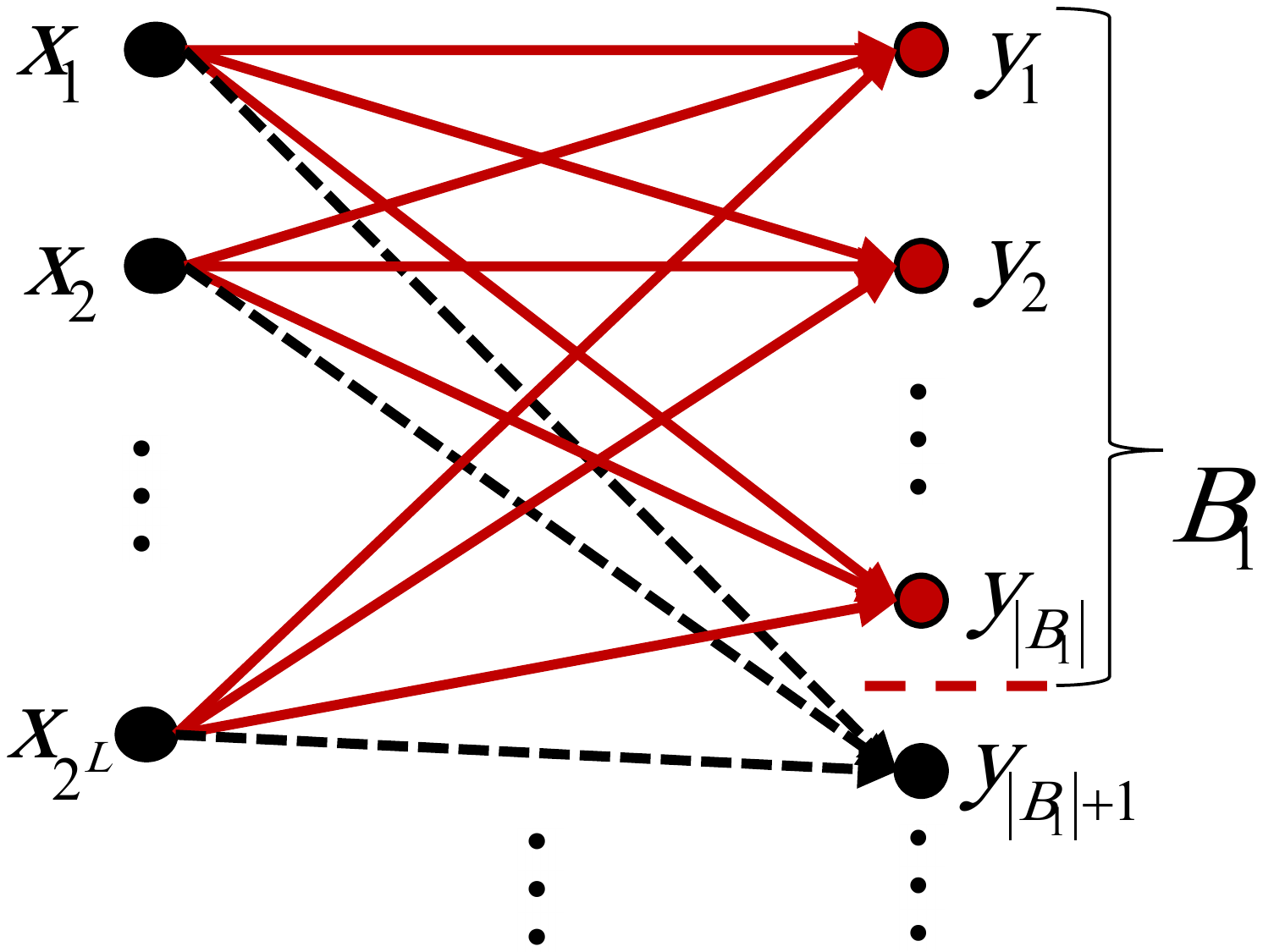}
		\caption{Partitioning of the DMC outputs.}
		\label{fig0}
	\end{figure}
	With the side information at the receiver, the $r$-th block of the auxiliary channel with input $X^L$ and output $Y^{L^\prime}$, where $Z^K$ denotes sequence $Z$ of length $K$, can be regarded as a DMC denoted as $Ch_{\lambda}\left(L\right)$ with $2^L$ inputs but an infinite number of outputs, due to Poisson-distributed repetitions, which makes the numerical calculation of the capacity infeasible. To resolve this issue, we first state the following lemma which demonstrates the relationship between the information rate of a DMC and some auxiliary channels with non-overlapping output partitions. An example of DMC output partitioning is depicted in Fig. \ref{fig0}, where $\mathcal{B}_1$ and $\left|\mathcal{B}_1\right|$ denote the first partition and the number of its corresponding outputs, respectively, with the elements belonging to the first partition shown in red (solid lines). 
	\begin{lemma} 
		(DMC output partitioning) Let $Ch\left(L\right)$ and $\mathcal{B}_k$'s be a DMC with input length $L$ and the non-overlapping partitions of its outputs (as depicted in Fig. \ref{fig0}), respectively. We define $Ch\left(L,\mathcal{B}_k\right)$ as $Ch\left(L\right)$ with outputs limited to $\mathcal{B}_k$ (and edges with normalized probabilities).
		Then, for a fixed input distribution $P\left(X^L\right)$, the relationship between the information rates of $Ch\left(L\right)$ and $Ch\left(L,\mathcal{B}_k\right)$'s can be written as
		\begin{equation}
			I\left(X^L;Y^{L^\prime}\right)  = \sum_k P\left({\mathcal{B}_k}  \right) R\left(Ch\left(L,\mathcal{B}_k\right),P\left(X^L\right)\right),
			\label{eq10b}
		\end{equation}
		where $I\left(\cdot;\cdot\right)$ denotes the mutual information, $P\left({\mathcal{B}_k}  \right)$ is the probability that the output of $Ch\left(L\right)$ belongs to the set ${\mathcal{B}_k}$, and $R\left(Z,P\right)$ is the mutual information of a DMC $Z$ under the input distribution $P$.
		\label{lemma1}
	\end{lemma}
	\begin{IEEEproof}
		When the side information is available only at the receiver, for a given input distribution, the mutual information for each block of the auxiliary channel can be written as:
		\begin{align}
				\hspace*{-5pt}I\Big(X^L;&Y^{L^\prime}\Big) \nonumber\\
				=& H\left(Y^{L^\prime}\right) - H\left(Y^{L^\prime} | X^L \right) \nonumber\\
				=& H\left(Y^{L^\prime}\right) \!-\! \sum_{j=1}^{2^L}  P\left(X^L=x_j\right) H\left(Y^{L^\prime} | X^L=x_j \right) \nonumber\\
				=& \!-\! \sum_{i=1}^{\infty} p_i \log_2p_i \!+\!  \sum_{j=1}^{2^L} \! P\left(X^L=x_j\right) \!\sum_{i=1}^{\infty}  p_{ji} \log_2  p_{ji}  , 
			\label{eq5}
		\end{align}
		where $p_{ji}=p\left(Y^{L^\prime}=y_i \left| X^L=x_j \right. \right)$ is the probability of receiving the output $y_i$ given the input $x_j$ (i.e., the $\left(i,j\right)$-th element of the transition matrix of $Ch\left(L\right)$), $p_i$ is the probability of receiving the output $y_i$, and $H\left(\cdot\right)$ denotes the entropy.
		
		For any arbitrary non-negative coefficient $c$, we have $\sum_{i} p_i \log_2p_i =\sum_{i} c\frac{p_i}{c} \log_2\left(c\frac{p_i}{c}\right) =c\sum_{i} \frac{p_i}{c} \log_2\left(\frac{p_i}{c}\right) +\log_2 c \sum_{i} p_i$.
		So, choosing $c = P\left({\mathcal{B}_k}  \right)$, we can decompose the mutual information in \eqref{eq5} as 
		\begin{align}
				\hspace*{-1cm}I\Big(X^L;&Y^{L^\prime}\Big) \nonumber\\
				=&  \!-\! \sum_k \! \!\sum_{i \in {\mathcal{B}_k}} \! p_i \log_2p_i \!+\! \sum_{j=1}^{2^L} \! P\left(x_j\right) \sum_k \!\!\sum_{i \in {\mathcal{B}_k}} \! p_{ji} \log_2  p_{ji} \nonumber\\
				=& -\sum_k P\left({\mathcal{B}_k}  \right)\sum_{i \in \mathcal{B}_k} \frac{p_i}{P\left({\mathcal{B}_k}  \right)} \log_2\left(\frac{p_i}{P\left({\mathcal{B}_k}  \right)}\right) \nonumber\\
				& - \sum_k \log_2 P\left({\mathcal{B}_k}  \right) \sum_{i \in {\mathcal{B}_k}} p_i  \nonumber\\
				& + \sum_{j=1}^{2^L}  \! P\left(x_j\right) \sum_k \! P\left({\mathcal{B}_k}  \right) \! \sum_{i \in {\mathcal{B}_k}} \! \frac{p_{ji}}{P\left({\mathcal{B}_k}  \right)} \log_2\left(\!\frac{p_{ji}}{P\left({\mathcal{B}_k}  \right)}\!\right) \nonumber\\
				& + \sum_{j=1}^{2^L}  P\left(x_j\right) \sum_k \log_2 P\left({\mathcal{B}_k}  \right) \sum_{i \in {\mathcal{B}_k}} p_{ji} \nonumber\\
				=& -\sum_k P\left({\mathcal{B}_k}  \right)  \left( \sum_{i \in \mathcal{B}_k} \frac{p_i}{P\left({\mathcal{B}_k}  \right)} \log_2\left(\frac{p_i}{P\left({\mathcal{B}_k}  \right)}\right) \right. \nonumber\\
				& \hspace{25pt}\left. - \sum_{j=1}^{2^L}  P\left(x_j\right) \sum_{i \in {\mathcal{B}_k}} \frac{p_{ji}}{P\left({\mathcal{B}_k}  \right)} \log_2\left(\frac{p_{ji}}{P\left({\mathcal{B}_k}  \right)}\right)\!\! \right) \nonumber\\
				=& \sum_k P\left({\mathcal{B}_k}  \right) R\left(Ch\left(L,\mathcal{B}_k\right),P\left(X^L\right)\right),
			\label{eq10}
		\end{align}
		where 
		\begin{align}
				\hspace*{-.23cm}R\big(Ch\left(L,\mathcal{B}_k\right),P&\left(X^L\right)\big) \nonumber\\
				=& -\sum_{i \in \mathcal{B}_k} \frac{p_i}{P\left({\mathcal{B}_k}  \right)} \log_2\left(\frac{p_i}{P\left({\mathcal{B}_k}  \right)}\right) \nonumber\nonumber\\
				&+ \sum_{j=1}^{2^L}  P\left(x_j\right) \sum_{i \in {\mathcal{B}_k}} \frac{p_{ji}}{P\left({\mathcal{B}_k}  \right)} \log_2\left(\frac{p_{ji}}{P\left({\mathcal{B}_k}  \right)}\right) 
		\end{align}
		is the mutual information of $Ch\left(L\right)$ with the output limitation $\mathcal{B}_k$.
	\end{IEEEproof}
	
	In the following, we employ Lemma \ref{lemma1} to obtain bounds on the capacity of a DMC with limited number of output bits (or equivalently, limited columns of the transition matrix), which can be numerically computed using the BAA. 
	Using Lemma \ref{lemma1}, Since $R\left(Z,\cdot\right)\le L$ for any DMC $Z$ with input length $L$, we can write lower and upper bounds on the input-output mutual information as
	\begin{equation}
			I\!\left(X^L;Y^{L^\prime}\right) \!\!\ge P\left({\mathcal{B}_1}  \right) R\left(Ch_{\lambda}\!\left(L,\mathcal{B}_1\right),\!P\left(X^L\right)\!\right),
		\label{I_bound1}
	\end{equation}
	and
		\begin{equation}
			I\!\left(X^L;Y^{L^\prime}\right) \!\!\le P\left({\mathcal{B}_1}  \right) R\left(Ch_{\lambda}\!\left(L,\mathcal{B}_1\right),\!P\left(X^L\right)\!\right) 
			\!+\! \left( 1 \!-\! P\left({\mathcal{B}_1}  \right) \!\right) L,
		\label{I_bound2}
	\end{equation}
	respectively, where ${\mathcal{B}_1}$ is the set of desired outputs (i.e., those that satisfy the length constraint). 
	Hence using the distribution $P\left(X^L\right)$ that maximizes the right-hand side (RHS) of \eqref{I_bound1}, and noting that the capacity of the $X^L \rightarrow Y^{L^\prime}$ channel $C_V$ is greater or equal to $I\left(X^L;Y^{L^\prime}\right)$ with that input distribution, we can write
		\begin{equation}
			C_V \ge \frac{1}{L} P\left({\mathcal{B}_1}  \right) f\left(Ch_{\lambda}\left(L,\mathcal{B}_1\right)\right),
		\label{cap_ineq1}
	\end{equation}
	where $C_V$ is the capacity of the auxiliary channel, and $f\left(Z\right) = \max_{P\left(X^L\right)} R\left(Z,P\left(X^L\right)\right)$ is the capacity of DMC $Z$, which can be numerically calculated using the BAA.

	For the upper bound, using the input distribution $P^{\prime}\left(X^L\right)$ that maximizes the left-hand side (LHS) of \eqref{I_bound2}, we have
	\begin{equation}
		C_V \le \frac{1}{L}P\left({\mathcal{B}_1}  \right)  R\left(Ch_{\lambda}\left(L,\mathcal{B}_1\right),P\left(X^L\right)\right) 
		+  1 - P\left({\mathcal{B}_1}  \right) ,
		\label{new5}
	\end{equation}
	where the RHS is computed under the input distribution $P^{\prime}\left(X^L\right)$. Using the input distribution that maximizes the RHS, the upper bound would still hold, hence we obtain
	\begin{equation}
	C_V \le \frac{1}{L} P\left({\mathcal{B}_1}  \right) f\left(Ch_{\lambda}\left(L,\mathcal{B}_1\right)\right) +  1 - P\left({\mathcal{B}_1}  \right).
	\label{cap_ineq}
\end{equation}

	
	\subsection{An Alternative Explanation}
	\begin{figure}
		\centering
		\includegraphics[trim={0cm 5cm 0cm 5cm},clip,width=1\linewidth]{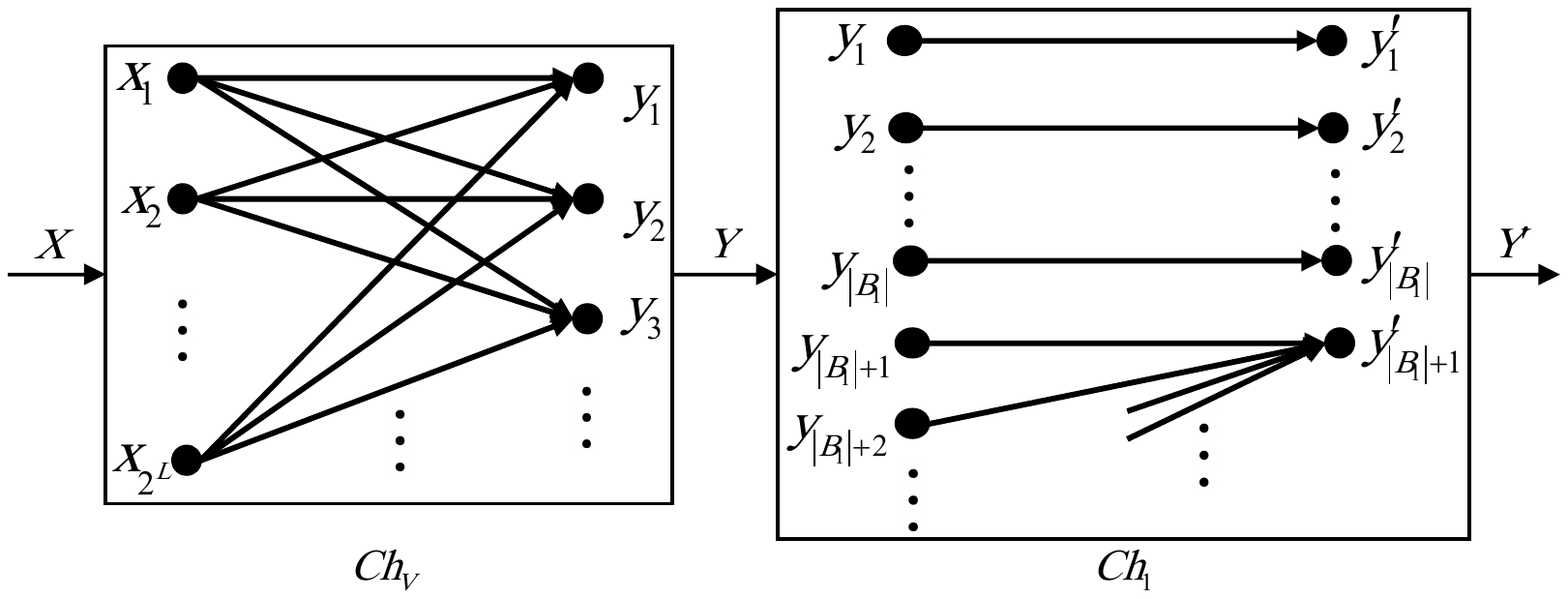}
		\caption{Concatenation of the auxiliary channel with another one.}
		\label{fig0b}
	\end{figure}
	
	We can think of the process of keeping only the outputs with smaller lengths than a threshold by concatenation of the auxiliary channel with another channel that passes the desired outputs as they are but combines the rest into a single output, as depicted in Fig. \ref{fig0b}. Using the data processing inequality \cite{Cover2006}, the capacity of the auxiliary ${Ch}_V$ channel is lower bounded by the capacity of the overall channel, which results in the lower bound in \eqref{cap_ineq}.
	
	Let us write the transition matrix of the auxiliary channel $P_V$ as the concatenation of two sets of columns as $P_V = \left[ P_V^{des} | P_V^{out}\right]$, with $P_V^{des}$ and $P_V^{out}$ being the columns corresponding to the desired outputs and the remaining ones, respectively. Now, we can represent the auxiliary channel as a concatenation of two channels with transition matrices 
	$\left[ P_V^{des} | P_{out} I_{2^L}\right]$ 
	and 
	$\left[\begin{tabular}{c:c}
		$I_{2^L}$ &  ${\bf 0}$\\
		\hdashline
		${\bf 0}$ & $P^{-1}_{out}P_V^{out}$\\
	\end{tabular}\right]$,
	respectively, where $I_k$ is the $k \times k$ identity matrix, and $P_{out}=1-P\left({\mathcal{B}_1}\right)$ is the probability of undesired outputs. Using the data processing inequality, the capacity of the auxiliary channel is upper bounded by the capacity of the first channel, as also given with the upper bound in \eqref{cap_ineq}. 
	
	\subsection{Upper Bound on the PRC Capacity}
	We know that the capacity of the auxiliary channel $C_V$ is an upper bound on that of the PRC $C_{\lambda}$. We set $\mathcal{B}_1$ as the set of outputs with the length $R$ smaller than a maximum output length $R_{m}$. Now, knowing that the output of each block is a Poisson random variable with parameter $L \lambda$ and using \eqref{cap_ineq}, the capacity of PRC $C_{\lambda}$ can be upper bounded as
		\begin{IEEEeqnarray*}{rCl}
			C_{\lambda} &\le& C_V\\
			&\le&
			\frac{1}{L}P\left( R \!\le\! R_{m} \right)  f\left(Ch_{\lambda}\left(L, R \!\le\! R_{m}\right)\!\right) 
			\!+\! 1 \!-\! P\left( R \!\le\! R_{m} \right)\\
			&=&\frac{1}{L} F\left(R_{m}; \! L \lambda\right)  f\left(Ch_{\lambda}\left(L, R \!\le\! R_{m}\right)\!\right) 
			\!+\! 1 \!-\! F\left(R_{m};\! L \lambda\right), \\
			 \IEEEyesnumber
		\label{eq9}
	\end{IEEEeqnarray*}
	where $F\left(R;\lambda\right)=\frac{\Gamma\left(R+1,\lambda \right)}{R ! }$ is the CDF of Poisson distribution with parameter $\lambda$ with  $\Gamma\left(\cdot,\cdot \right)$ denoting the upper incomplete gamma function.

	\section{Bounds with Limitation on the Number of Repetitions of Each Bit}
	The computational complexity of generating the corresponding transition matrix and calculating the proposed upper bound grows exponentially with the input length $L$ and polynomially with the output length $R$ (or $R_{m}$). This makes obtaining numerical results for large $L$ and $R$ values infeasible, both in terms of the run-time and memory requirements. 
	To alleviate this issue to some extent, we limit the number of repetitions per input bit, $\bar R_{m}$, as well, which is especially beneficial for low values of $\lambda$ since large numbers of repetitions are very unlikely.
	However, this limitation cannot be addressed by DMC output partitioning as in Lemma \ref{lemma1} since there is no one-to-one correspondence between a per-bit limitation and any specific set of outputs. For instance, let $L=2$ and the per-bit limitation be $\bar R_{m}\le1$. In this case, inputs `01' and `11' with repetition patterns $(0,2)$ and $(1,1)$, respectively, lead to the same output `11', where the first pattern violates the per-bit limitation while the second one does not. 
	
	To resolve this issue, we first define side information $S$ at the receiver, with $S=1$ if the condition on the number of bit repetitions and output length is satisfied, and $S=0$ otherwise. For any input distribution, we can write
		\begin{IEEEeqnarray*}{rCl}
	I\left(X^L;Y^{L^\prime},S\right)  &=& I\left(X^L;Y^{L^\prime}\right)
	+I\left(X^L;S|Y^{L^\prime}\right)\\
	&=& I\left(X^L;S\right) + I\left(X^L;Y^{L^\prime}|S\right). \IEEEyesnumber
		\label{new1}
	\end{IEEEeqnarray*}
	
	Since the repetition patterns and the channel input are independent, $S$ and $X^L$ are independent, and hence $I\left(X^L;S\right)=0$. Therefore, using \eqref{new1} and the fact that $I\left(X^L;S|Y^{L^\prime}\right) \le H\left(S\right)$, for any input distribution, we can write
		\begin{IEEEeqnarray*}{rCl}
			I\left(X^L;Y^{L^\prime}\right) &=& I\left(X^L;Y^{L^\prime}|S\right) - I\left(X^L;S|Y^{L^\prime}\right)\\
			&\ge&  I\left(X^L;Y^{L^\prime}|S\right) - H\left(S\right)\\
			&\ge&  P_s I\left(X^L;Y^{L^\prime}|S=1\right) - H\left(S\right), \IEEEyesnumber
		\label{new2}
	\end{IEEEeqnarray*}
	where $P_s=P\left(S=1\right)$. Using similar arguments as in the capacity lower bound in Section III.A, we have
	\begin{equation}
		C_V \ge P_s \left(\max_{P\left(X^L\right)} I\left(X^L;Y^{L^\prime}|S=1\right)\right) - H\left(S\right).
		\label{new3}
	\end{equation}
	Note that by selecting sufficiently large number of repetition patterns, $P_s$ can be made very close to 1, i.e., $H\left(S\right)$ can be made near zero.
	
	To obtain an upper bound on the channel capacity, we write
		\begin{IEEEeqnarray*}{rCl}
			\hspace{-10pt} I\left(X^L;Y^{L^\prime}\right) &\le & I\left(X^L;Y^{L^\prime}|S\right) \\
			&=&P_s I\left(X^L;Y^{L^\prime}|S=1\right)\\
			&&+\left(1-P_s\right)I\left(X^L;Y^{L^\prime}|S=0\right)\\
			&\le& P_s I\left(X^L;Y^{L^\prime}|S=1\right)
			+\left(1-P_s\right)L, \IEEEyesnumber
		\label{new4}
	\end{IEEEeqnarray*}
	where the last inequality is obtained by bounding $I\left(X^L;Y^{L^\prime}|S=0\right)$ trivially by $L$.
	
	Using the input distribution $P^{\prime}\left(X^L\right)$ that maximizes the LHS of \eqref{new4}, we have
\begin{equation}
	C_V \le P_s I\left(X^L;Y^{L^\prime}|S=1\right)
	+\left(1-P_s\right)L,
	\label{new60}
\end{equation}
where the RHS is also computed under $P^{\prime}\left(X^L\right)$. Noting that the upper bound would still hold if we use the input distribution that maximizes the RHS, we obtain
		\begin{equation}
		C_V \le P_s \left(\max_{P\left(X^L\right)} I\left(X^L;Y^{L^\prime}|S=1\right)\right) +\left(1-P_s\right)L.
		\label{new6}
	\end{equation}

	\section{On Lower Bounding on the PRC Capacity} \label{Rx.LB}
	In this section, we offer a way of developing lower bounds on the capacity of PRC.
	To this end, similar to the idea in \cite{Fertonani2010}, we write
		\begin{IEEEeqnarray*}{rCl}
			I\left(X^N;Y^{N^\prime}\right) &=& I\left(X^N;Y^{N^\prime},V^Q\right) - I\left(X^N;V^Q|Y^{N^\prime}\right)\\
			&\ge& I\left(X^N;Y^{N^\prime},V^Q\right) - H\left(V^Q|Y^{N^\prime}\right)\\
			&=& I\left(X^N;Y^{N^\prime},V^Q\right) + H\left(Y^{N^\prime}\right) \\
			&&- H\left(Y^{N^\prime}|V^Q\right) - H\left(V^Q\right), \IEEEyesnumber
		\label{LB1}
	\end{IEEEeqnarray*}
	where $N^\prime$ is the length of the channel output sequence.
	Letting input length $N$ go to infinity, for any input distribution, we can write
		\begin{IEEEeqnarray*}{rCl}
			C_{\lambda} \!&\ge&  \lim_{N \to \infty} \frac{1}{N} I\left(X^N;Y^{N^\prime}\right)\\
			&\ge&  \! \lim_{N \to \infty} \frac{1}{N} \left(I\left(X^{\!N};Y^{\!N},\! V^{\!N}\right) \!+\! H \! \left(Y^{\!N}\! \right) \!-\! H \! \left(Y^{\!N}|V^{\!N}\right)\!\right) \\
			&&- \lim_{N \to \infty} \frac{1}{N} H \left(V^{\!N}\right)\\
			&\ge& \frac{1}{L} I\left(X^L;Y^{L^\prime}\right) \\
			&&- \! \lim_{N \to \infty} \!\frac{1}{N} \! \left( \! H \left(V^Q\right) \!+\! H\left(Y^{N^\prime}\right) \!-\! H\left(Y^{N^\prime}\!|V^Q\right)\!\! \right) . \IEEEyesnumber
	\end{IEEEeqnarray*}
	Since this is true for any input distribution, as a reasonable choice one can pick the one that maximizes the first term on the RHS.
	
	Also, for the entropy of the side information, we have
		\begin{IEEEeqnarray*}{rCl}
			\lim_{N\to\infty} \frac{1}{N} H\left(V^Q\right) &=& 
			\lim_{N\to\infty} \frac{1}{N} H\left(V_1,V_2,\cdots,V_Q\right)\\
			&=& \frac{1}{L}\lim_{Q\to\infty} \frac{1}{Q} H\left(V_1,V_2,\cdots,V_Q\right)\\
			&=& \frac{1}{L}\lim_{Q\to\infty} \frac{1}{Q} \sum_{i=1}^Q H\left(V_i\right)\\
			&=& \frac{1}{L} H\left(V_i\right), \IEEEyesnumber
		\label{Hv}
	\end{IEEEeqnarray*}
	since $V_i$'s are independent and identically distributed (i.i.d.) random variables. 
	
	For the conditional entropy, we can write
		\begin{IEEEeqnarray*}{rCl}
			\lim_{N\to\infty} \! \frac{1}{N} H \!  \left(Y^{\!N} \! |V^{\!N}\right) \!&=& \! \lim_{N\to\infty} \!  \frac{1}{N} H\left(Y_1,\! Y_2,\cdots \! ,\! Y_Q|V_1,\! V_2,\cdots \! ,\! V_Q\right) \\
			&\le& \lim_{N\to\infty} \frac{1}{N} \sum_{i=1}^Q H\left(Y_i|V_1,V_2,\cdots,V_Q\right)\\
			&=& \lim_{N\to\infty} \frac{1}{N} \sum_{i=1}^Q H\left(Y_i|V_i\right)\\
			&=& \frac{1}{L} \lim_{Q\to\infty} \frac{1}{Q} \sum_{i=1}^Q H\left(Y_i|V_i\right)\\
			&=& \frac{1}{L} H\left(Y_i|V_i\right)\\
			&=& \frac{1}{L} \sum_{v=0}^{\infty} P\left(V_i=v\right) H\left(Y_i|V_i=v\right), \IEEEyesnumber
		\label{cond_ent}
	\end{IEEEeqnarray*}
	since $H\left(X_1,\cdots,X_N\right)\le \sum_i H\left(X^L\right)$ and that $Y_j$ is independent of $V_i$'s when $i \neq j$.
	
	Combining these, we arrive at
		\begin{IEEEeqnarray*}{rCl}
		\hspace{-5pt}	C_{\lambda} &\ge& 
			C_V + \lim_{N \to \infty} \frac{1}{N}\left[H\left(Y^{N^\prime}\right)\right]_{P^*_1\left(X\right)}\\
			&&- \frac{1}{L} \left( H\left(V_i\right) + \sum_{v=0}^{\infty} P\left(V_i=v\right) H\left(Y_i|V_i=v\right)\right), \IEEEyesnumber
		\label{LB2}
	\end{IEEEeqnarray*}
	where ${P^*_1\left(X\right)}$ is obtained by repeating the input distribution that maximizes $I\left(X^L;Y^{L^\prime}\right)$ in all blocks. 
	
	Note that we cannot have a similar expression as in \eqref{Hv} for the output entropy $H\left(Y\right)$ since the output bits are not independent of each other. This is even the case when the input bits are independent. Since a direct calculation of $H\left(Y\right)$ does not seem feasible, it needs to be properly lower bounded, which needs a deeper analysis. A simple way to circumvent this issue is to use the fact that $H\left(Y\right) - H\left(Y|V\right) = I\left(Y;V\right)\ge0$. Hence, combining \eqref{LB1} and \eqref{LB2}, we obtain a looser but more easily calculable lower bound as 
	\begin{equation}
			C_{\lambda} \ge 
			C_V - \frac{1}{L} H\left(V_i\right),
		\label{LB3}
	\end{equation}
	which is in a similar form with the one on the BDC capacity obtained in \cite{Fertonani2010}. Further lower bounding the RHS by substituting the expression in \eqref{new3} into \eqref{LB3}, and using the per-bit and overall length limitations, we finally obtain
		\begin{IEEEeqnarray*}{rCl}
			C_{\lambda} &\ge& 
			\frac{1}{L}P\left(R_i \le \bar R_{m} \cap \ R \le R_{m} \right) \\
			&&\times f\left(Ch\left(L,R_i \le \bar R_{m} \cap R \le R_{m}\right)\right)  \\ 
			&&- \frac{1}{L} H\left(S\right) - \frac{1}{L} H\left(V_i\right).
			\IEEEyesnumber
		\label{LB4}
	\end{IEEEeqnarray*}
	
	Since the assumed side information $V_i$ is a Poisson random variable with parameter $L\lambda$, we can easily compute its entropy numerically, or resort to the approximations in \cite{Evans1988,Cheraghchi20192}. Or, as an alternative, we can employ the simple upper bound provided in \cite{Adell2010}, based on the generic upper bound in \cite[Theorem 8.6.5]{Cover2006}, as 
		$H\left(Z_{\lambda}\right)\le \frac{1}{2} \log_2 \left( 2 \pi e\left(\lambda + \frac{1}{12}\right)\right)$,
	where $Z_{\lambda}$ is a Poisson random variable with parameter $\lambda$.

	
	\section{Numerical Results}
	We compare the proposed upper and lower bounds with the existing ones in the literature.
	Note that these bounds are calculated by running the BAA until a certain precision is obtained.
	Let $f_{BAA}$ and $\epsilon_{BAA}$ be the capacity calculated via BAA and the precision of BAA, respectively. Since we know that $0\le f-f_{BAA} \le \epsilon_{BAA}$, the proposed lower bounds still hold using the BAA. However, for the upper bounds to hold, $\epsilon_{BAA}$ needs to be taken into account. So the modified bounds become
		\begin{IEEEeqnarray}{rCl}
			 C_{\lambda} \!&\le& \!\frac{1}{L}P\left(\mathcal{B}_1 \!\right) 
			\left(f_{BAA}\left(Ch\left(L,\mathcal{B}_1 \!\right)\!\right) \!+\! 
			\epsilon_{BAA} \right) 
			\!\!+\!\! 1 \!-\! P\left(\mathcal{B}_1 \!\right),\\
			C_{\lambda} \!&\ge& \!\frac{1}{L}P\left(\mathcal{B}_1 \right) 
			f_{BAA}\left(Ch\left(L,\mathcal{B}_1\right)\right)  
			\!-\! \frac{1}{L} H\left(V_i\right),
		\label{UB_LB}
	\end{IEEEeqnarray}
	which are the bounds we use in the numerical results with $\epsilon_{BAA} = 0.005$.
	
\begin{figure}
		\centering
		\includegraphics[trim={1.5cm 0 2cm 1cm},clip,width=1\linewidth]{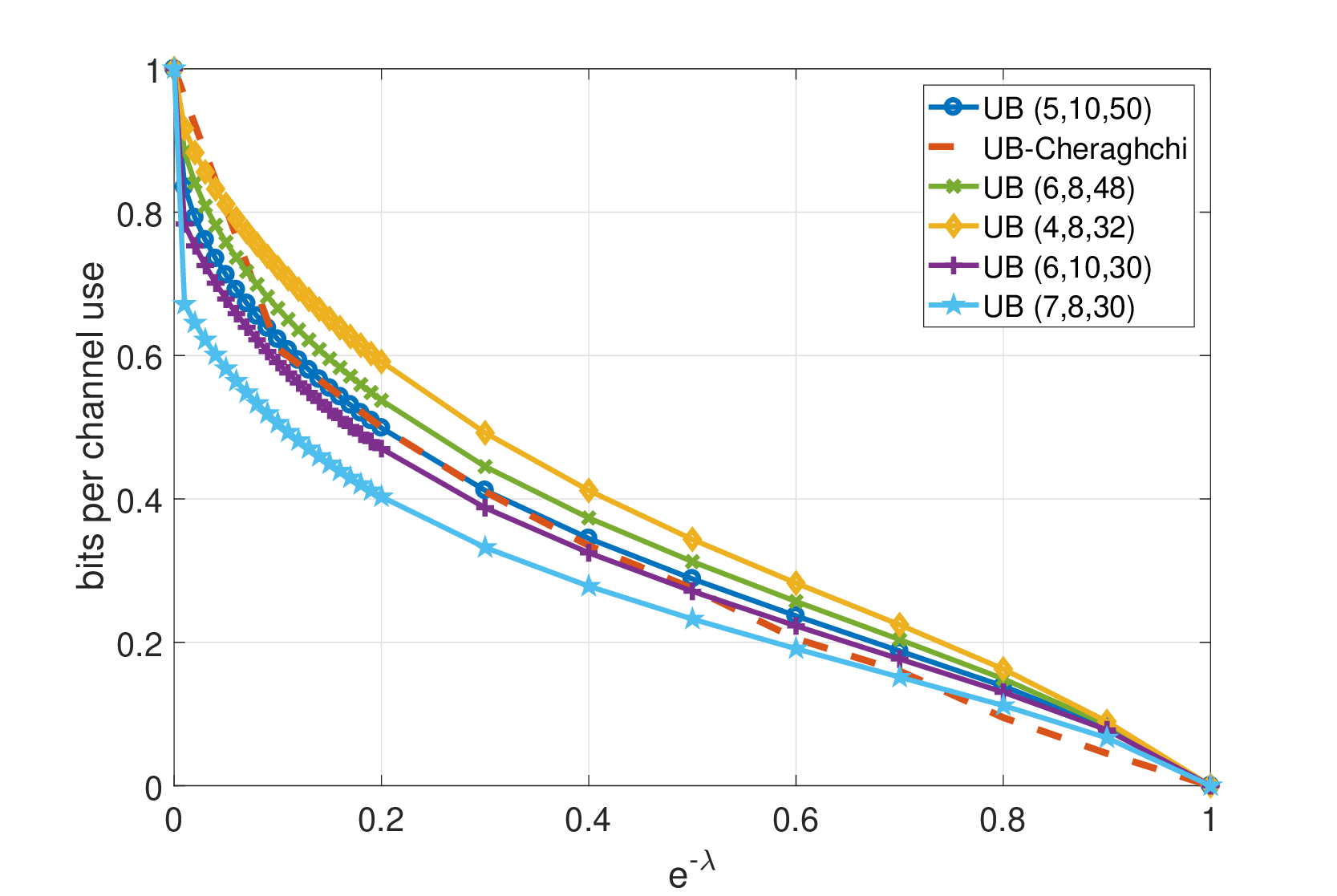}
		\caption{Upper bounds for different values of the parameter $\lambda$.}
		\label{fig1}
	\end{figure}

	
	The proposed upper bound for different $\left(L,\bar R_{m},R_{m}\right)$ are plotted in Fig. \ref{fig1}. “UB-Cheraghchi” is the tightest upper bound, provided in \cite{Cheraghchi2019}. The figure shows that the proposed upper bound improves the results in \cite{Cheraghchi2019} for $\lambda \ge 0.36$ ($e^{-\lambda}\le 0.7$) by up to 15\%.
	
	Note that with the set of parameters that we could utilize, up to $\left(L=7,\bar R_{m}=8,R_{m}=30\right)$ for large $R_{m}$'s and $\left(L=15,\bar R_{m}=2,R_{m}=7\right)$ for small $R_{m}$'s, due to computational limitations, we could not obtain any nontrivial (nonzero) result for the obtained (looser, but more easily calculable) lower bound in \eqref{UB_LB}. 
	However, it can be used as a basis for obtaining tighter bounds, e.g., by providing a tighter lower bound on $H\left(Y\right) - H\left(Y|V\right)$ (or, on $\lim_{N \to \infty} \frac{1}{N}\left[H\left(Y^{N^\prime}\right)\right]_{P^*_1\left(X\right)}$). This is left as future work.

	\section{Conclusion}
	We developed bounds on the capacity of the PRC. To this end, we first considered an auxiliary channel with the output lengths corresponding to input blocks of a certain length given as side information, which decomposes the channel into DMCs with a finite number of inputs. 
	Using some results on DMC output partitioning, and based on conditioning on an event about the repetition structure, we provided upper and lower bounds on the capacity of the auxiliary channel, which are in turn used to obtain bounds on the capacity of the PRC. Numerical evaluations reveal that the proposed upper bound is tighter than the best known one in the literature. 
	
	\section{Acknowledgments}
	This work was funded by the European Union ERC TRANCIDS 101054904. Views and opinions expressed are however those of the author(s) only and do not necessarily reflect those of the European Union or the European Research Council Executive Agency. Neither the European Union nor the granting authority can be held responsible for them.

\end{document}